\documentclass[a4,aps,prd,superscriptaddress,floatfix,nofootinbib,showpacs]
{revtex4}

\usepackage{color}
\usepackage[dvips]{graphicx}

\newcommand{\lsim}{\mathrel{\mathop{\kern 0pt \rlap
  {\raise.2ex\hbox{$<$}}}
  \lower.9ex\hbox{\kern-.190em $\sim$}}}
\newcommand{\gsim}{\mathrel{\mathop{\kern 0pt \rlap
  {\raise.2ex\hbox{$>$}}}
  \lower.9ex\hbox{\kern-.190em $\sim$}}}

\newcommand{\be}{\begin{equation}}
\newcommand{\ee}{\end{equation}}
\newcommand{\beqarr}{\begin{eqnarray}}
\newcommand{\eeqarr}{\end{eqnarray}}

\begin{document}


\title{Gravitational reheating in  quintessential inflation}



\author{E. J. Chun$^1$, S. Scopel$^2$, I. Zaballa}
\affiliation{Korea Institute for Advanced Study (KIAS)\\
Hoegiro 87, Dongdaemun-gu, Seoul 130-722, Korea\\
$^2$School of Physics and Astronomy, Seoul National University\\
Gwanakro 599, Gwangak-gu, Seoul 151-749, Korea}

\date{\today}

\begin{abstract}
We provide a detailed study of gravitational reheating in
quintessential inflation generalizing previous analyses only available
for the standard case when inflation is followed by an era dominated
by the energy density of radiation.  Quintessential inflation assumes
a common origin for inflation and the dark energy of the Universe. In
this scenario reheating can occur through gravitational particle
production during the inflation-kination transition. We calculate
numerically the amount of the radiation energy density, and determine
the temperature $T_*$ at which radiation starts dominating over
kination. The value of $T_*$ is controlled by the Hubble parameter
$H_0$ during inflation and the transition time $\Delta t$, scaling as
$H_0^2\, [\ln(1/H_0\Delta t)]^{3/4}$ for $H_0 \Delta t \ll1$ and
$H_0^2 (H_0 \Delta t)^{-c}$ for $H_0\Delta t \gg 1$.  The
model-dependent parameter $c$ is found to be around 0.5 in two
different parametrizations for the transition between inflation and
kination.
\end{abstract}


\maketitle

\section{Introduction}
\label{sec:intro}

Over the last few years strong evidence has been found allowing to
conclude that the Universe underwent an accelerated expansion with
negative pressure for at least twice in its history: in its very early
infancy during inflation, and at the present day. While inflation was
introduced as an elegant solution to the horizon and flatness problem
and to explain the anisotropy of the cosmic microwave background
radiation \cite{inflation}, today's accelerated expansion was an
unexpected discovery which strongly favors the existence of a dark
energy component contributing a fraction $\Omega_{\rm DE}\simeq$ 0.7
to the closure density. It is somehow intriguing that both periods of
accelerated expansion can be explained by scalar fields whose
potential energy dominates the energy density at some time of its
evolution. In fact, while for inflation an ``inflaton'' field is
introduced, the dark energy component in the present Universe can be
explained in a dynamical way by modifying the standard cosmology with
the introduction of a slowly evolving scalar field called
``quintessence'' \cite{caldwell}.  Compared to a cosmological
constant, the latter approach has the nice feature of explaining in a
natural way why radiation and dark energy provide comparable
contributions to the energy budget of the present Universe, in spite
of having very different time evolutions (the coincidence problem),
through ``tracking solutions'' \cite{tracking} of the quintessence
field. It is then natural to imagine whether it is possible to unify
the two scenarios, identifying inflaton and quintessence with the same
field $\phi$ \cite{spokoiny,peebles}.  As a consequence of this,
various models of quintessential inflation have been discussed in the
literature \cite{quintessential_inflation}.

In the quintessential inflation scenario two main qualitative
properties emerge as follows.

i) The potential $V(\phi)$ needs to account for the large mismatch
between the inflationary plateau at the beginning of the $\phi$ field
evolution (whose natural value $V_0$ is expected to be within a few
orders of magnitude of the Plank scale $m_P=2.4\times10^{18}$ GeV) and
the tiny scale of the quintessential tail $V_F$ eventually accounting
for the cosmological constant today, about $10^{-121}$ times
smaller. As a consequence of this, a crucial requirement of $V(\phi)$
is to have a rapid fall at the end of inflation, so that at the end of
its slow--roll phase the field $\phi$ experiences a strong
acceleration as it ``deep-dives'' from $V_0$ toward $V_F$.  Thus the
Universe undergoes a period of ``kination'' expansion when its energy
density is dominated by the kinetic energy of $\phi$, whose potential
energy eventually provides the dark energy at later time.

ii) In this scenario the standard reheating mechanism which is usually
assumed to create the initial plasma by the decay of $\phi$ is not at
work.  However, the mechanism of gravitational particle production  is
able to reheat the Universe \cite{ford,starobinsky_birrel}
 without introducing extra ingredients in
the scenario of quintessential inflation.  Although it is much less efficient
than the usual reheating mechanism, there is no difficulty to
accommodate the required cosmology of radiation domination.  Initially
at the gravitational reheating temperature, the tiny amount of
radiation produced by this mechanism is largely sub-dominant compared
to the energy contribution from kination. However, since the energy
density of kination is red-shifted away by the cosmological expansion
much faster than the radiation density, radiation eventually dominates
at some lower temperature $T_*$ which, in viable models, must be above
the MeV scale in order to preserve the successes of Big Bang
Nucleosynthesis.  Here we remark that other reheating mechanisms like
preheating \cite{felder} or curvaton reheating \cite{feng} could be
realized typically by introducing additional scalar fields in
quintessential inflation.

\medskip

The purpose of this work is to provide a quantitative analysis for the
amount of particle production and determine the temperature $T_*$
assuming the gravitational reheating mechanism.  The actual value of
$T_*$ has important phenomenological consequences. During kination the
expansion rate of the Universe is larger compared to the usual
radiation domination case, and this can lead to various non-standard
cosmological scenarios.  For instance, if $T_*$ is at the GeV scale
the relic abundance of a thermal cold dark matter candidate can be
significantly enhanced compared to the canonical prediction
\cite{salati}, because its decoupling time from the plasma is
anticipated, providing the correct amount of Dark Matter for values of
the cross sections between the DM particle and ordinary matter
sizeably higher compared to the standard case. This has interesting
phenomenological implications not only for the LHC and other future
collider experiments but also for astrophysical observations like
direct or indirect dark matter detection.  Indeed, the enhanced dark
matter annihilation rate which is hinted by the recent PAMELA
\cite{PAMELA} or ATIC/PPB-BETS \cite{PPB-BETS,ATIC} data, can be made
compatible with the thermal dark matter production in the context of
quintessential inflation.  The presence of a kination era may also
have an impact on the properties of the electroweak baryogenesis
\cite{joyce} or the thermal leptogenesis induced by the CP-violating
decay of a right-handed neutrino \cite{chun_scopel}.

Due to its interesting phenomenological implications, we aim to
discuss $T_*$ in detail within the quintessential inflation
scenario. The original calculation of gravitational reheating has
been performed in Ref.~\cite{ford} in the case of the
inflation-radiation transition, and its results have been widely
used also in papers discussing quintessential inflation
\cite{quintessential_inflation}.  In our discussion we will
generalize the analysis of Ref.~\cite{ford} to the case of the
inflation-kination transition by performing full numerical
calculations of the Bogolubov coefficients as is reviewed in
Section 2.  In Section 3 we introduce two ways of parametrizing
the transitions between inflation and radiation or kination, in
order to discuss how the particular way of modeling them may
affect our conclusions.  The results of our numerical calculation
of the radiation energy density are presented in Section 4 and the
kination-radiation equality temperature $T_*$ is discussed in
Section 5, showing its dependence on the inflationary scale and
the transition time.  We give our conclusions in Section 6.

\section{Gravitational particle creation: A setup}
\label{sec:reheating}

We briefly review here the mechanism of gravitational particle
creation \cite{birrell,Fordreview} in order to set up the relevant
equations in a suitable form for numerical integration.  The main
results of this section are Eqs.~(\ref{eq:final_eq}) and
(\ref{eq:final_beta}). In a spatially flat background the
Robertson-Walker metric is written as:

\be d
s^2=a^2(\eta)(d\eta^2-dx^idx^i), \label{eq:rw} \ee

where $a(\eta)$ is the scale factor as a function of the conformal
time $\eta$.  Recall that the conformal time $\eta$ is related to
the usual time $t$ by $d\eta=dt/a$.  Let us consider a massless
scalar field $\Phi$ minimally coupled to gravity. Spatial
translation symmetry allows to separate spatial and time
dependence in the Fourier-transform of the field:
$\phi_k(\vec{x},t)=\rho_k(t) e^{i\vec{k}\cdot \vec{x}}
\label{eq:states}$. Factoring out the scale factor: $\chi_k \equiv
\rho_k/a$,  one finds  the equation of motion of the (conformal)
time--dependent harmonic oscillator:
\begin{eqnarray}
&&\chi_k^{\prime\prime}+\omega_k^2(\eta)\chi_k=0
\label{eq:harm_osc} \\
\mbox{where}&&
\omega_k^2(\eta)=k^2+V=k^2-\frac{a^{\prime\prime}}{a}.
\label{eq:omega_k}
\end{eqnarray}
Here the prime indicates a derivative with respect to conformal
time and $k$ is the comoving momentum. Assuming the adiabatic
condition $V \ll k^2$ in the early past and in the late future (so
that $\omega(\eta\rightarrow \pm \infty)$=constant) implies that
the field operator can be expanded in terms of a complete set of
positive--frequency solutions. Denoting  the solutions
asymptotically free in the past and in the future by ${f_j}$ and
${F_j}$, respectively, we have
 \be
 \phi=\sum_j\left(a_j
f_j+a^{\dagger}_j f^*_j \right)=\sum_j\left(b_j F_j+b^{\dagger}_j
F^*_j \right). \label{eq:bog1}
 \ee
In the above equation $a_j$ and $b_j$ are operators that
annihilate the {\it in} and {\it out} vacua, and are connected by
the Bogolubov transformations:
 \beqarr
a_j&=&\sum_k\left (\alpha^*_{jk} b_k-\beta^*_{jk}b^{\dagger}_k
\right)\nonumber \\
b_k&=&\sum_j\left (\alpha_{jk} a_j+\beta^*_{jk}a^{\dagger}_j
\right). \label{eq:bogoliubov_trans}
 \eeqarr
Assuming that at $\eta=-\infty$ no particles are present, the
initial vacuum $|0\rangle_{in}$, which in the Heisenberg picture
is the state of the system for all the time, is annihilated by the
$a_j$ operators: $a_j|0\rangle_{in}$=0. However the physical
number operator that counts particles in the out-region is
$N_k=b^{\dagger}_k b_k$, so that the mean number of particles
created into mode $k$ is $ \langle N_k \rangle=_{in}\langle
0|b^{\dagger}_k b_k|0 \rangle_{in}=\sum_j|\beta_{jk}|^2
\label{eq:n_k}$, which implies that particle creation is
proportional to the negative--frequency content of {\it out}
states with the boundary condition that the {\it in} states
contain only negative frequencies. This is tantamount to solving
Eq.(\ref{eq:harm_osc}) with the boundary conditions:
 \beqarr
\chi_k(\eta\rightarrow -\infty)&=&\frac{e^{-i k \eta}}{\sqrt{2k}}
\nonumber \\
\chi_k(\eta\rightarrow +\infty)&=&\frac{1}{\sqrt{2k}}\left
(\alpha_k e^{-i k \eta}+ \beta_k e^{i k \eta}\right ),
\label{eq:boundary}
 \eeqarr
\noindent with $k\equiv |\vec{k}|$.
The Wronskian condition: $\chi_k
\chi_k^{*\prime}-\chi_k^*\chi_k^{\prime}=i$
keeps the correct normalization of states
and implies the additional relation:
 \be|\alpha_k|^2-|\beta_k|^2=1.
\label{eq:alpha_meno_beta}
 \ee

Once the function $\omega(\eta)$ is given, setting
$\chi_k=r_k e^{i\phi_k}$
\noindent the real and imaginary part of Eq.~({\ref{eq:harm_osc}})
can be separated: \beqarr
&&r_k^{\prime\prime}+(\omega_k^2-\Omega_k^2)r_k=0 \nonumber\\
&&r_k \Omega_k^{\prime}+2 r_k^{\prime}\Omega_k=0.
\label{eq:diff_eq} \eeqarr where $\Omega_k\equiv \phi_k^{\prime}$.
The corresponding boundary conditions are
 \beqarr
&&r(\eta\rightarrow -\infty)=\frac{1}{\sqrt{2k}} \nonumber \\
&&r^{\prime}(\eta\rightarrow -\infty)=0 \nonumber\\
&&\Omega_k(\eta\rightarrow -\infty)=-k \label{eq:boundary_r}\,.
 \eeqarr
The last equation of (\ref{eq:diff_eq}) is separable and readily
solved, yielding
$\Omega_k=-\frac{1}{2 r_k^2}$.
Then the problem reduces to solving the single second-order
differential equation:
 \be r_k^{\prime\prime}+\left
(\omega_k^2(\eta)-\frac{1}{4 r_k^4}\right )r_k=0,
\label{eq:final_eq}
 \ee
whose solutions can be found by numerical integration given the
function $\omega_k(\eta)$ from Eq.(\ref{eq:omega_k}). Once the
values of $r_k(\eta\rightarrow \infty)$ and
$r_k^{\prime}(\eta\rightarrow \infty)$ are found, the Bogolubov
coefficient can be extracted using a suitable combination of
$\chi_k$ and its derivative $\chi_k^{\prime}$. Making use of the
asymptotic conditions (\ref{eq:boundary}) and of Eq.
(\ref{eq:alpha_meno_beta}), one finds \be
|\beta_k|^2=\frac{k^2|\chi_k|^2+|\chi_k^{\prime}|^2-k}{2
k}=\frac{k^2
  r_k^2+(r_k^{\prime})^2+\frac{1}{4 r_k^2}-k}{2 k}.
\label{eq:final_beta} \ee In the above equation for
$\eta\rightarrow -\infty$ the boundary conditions $r=1/\sqrt{2k}$,
$r^{\prime}=0$ consistently imply $|\beta_k|^2$=0. Note that
whenever $V \ll k^2$ (i.e. close to the adiabatic regime) a
perturbative approach is possible in the calculation of the
Bogolubov coefficient $\beta_k$, which can be approximated by the
Fourier transform of the function V \cite{starobinsky_birrel}: \be
\beta_k\simeq -\frac{i}{2
  k}\int_{-\infty}^{+\infty} e^{-2 i k \eta}
V(\eta)\,d\eta. \label{eq:fourier} \ee Although we will mainly
rely on the numerical solutions of Eq.(\ref{eq:final_eq}) for our
discussions, we will also give some examples where the numerical
solution is compared to Eq.(\ref{eq:fourier}). Given the Bogolubov
coefficient $\beta_k$, one obtains directly the final energy
density produced: \be \rho=\frac{1}{(2\pi a)^3 a}\int d^3\vec{k} k
|\beta_k|^2. \label{eq:energy_density} \ee

\section{Gravitational reheating for inflation--radiation/kination transition}
\label{sec:transition}

The amount of gravitational particle production can be calculated by
solving numerically Eq.~(\ref{eq:final_eq}) given the potential
$V=-a^{''}/a=-a^2R/6$ in $\omega_k(\eta)$ of Eq.~(\ref{eq:omega_k})
(here $R$ is the Ricci scalar).  Its functional forms are uniquely
defined during the inflation and radiation/kination periods, while a
certain degree of arbitrariness arises from the modelization of the
transition period.  For this reason in our analysis we will adopt two
different parametrizations for the transition.  One is a polynomial
parametrization for $a^2(\eta)$ following the original paper
\cite{ford}, and the other is a parametrization of the equation state
$w\equiv p/\rho$ smoothly connecting the inflation ($w=-1$) and the
radiation ($w=1/3$) or kination ($w=1$) regimes with an hyperbolic
tangent.

Let us first consider the ``standard'' situation where an inflationary epoch is
followed by a radiation--dominated period, which has been first discussed
in Ref.~\cite{ford}.
 Taking the (arbitrary) boundary condition,
$a(-H_0^{-1})=1$, with the conformal time normalized as
$\eta=-H_0^{-1}$ at the end of inflation, one gets during
inflation $a(t)=e^{H_0 t}$:
 \be
a(\eta)=-\frac{1}{H_0\eta}=-\frac{1}{x},
\label{eq:a_eta_inflation}
 \ee
where we have set $x\equiv
H_0\eta$. Recalling that the Ricci scalar is given by
$R=3(1-3w)H^2$, one has $R=12 H_0^2$ during inflation. If the universe is
dominated by radiation after inflation, the adiabatic condition is
automatically verified since $R=0$ with $w=1/3$, and  one has
$a^{\prime\prime}=0$ which implies that $a$ needs to be a linear
function of $\eta$.
Note that, assuming an abrupt transition at some time $\eta_0$: $R=12
H_0^2$ and $R= 0$ for $\eta<\eta_0$ and $\eta>\eta_0$ respectively,
one finds an ultra-violet divergence in the energy density of
gravitationally produced particles because the discontinuous change in
the metric produces too many particles in the high frequency modes
\cite{ford}. This is just an artifact of the simplified situation of
an abrupt transition, neglecting the fact that the transition takes
place during a finite interval in the conformal time $\Delta\eta$.  It
is then convenient to introduce the parameter $x_0=H_0\Delta\eta$, and
to parametrize the function $a^2(\eta) \equiv f(x)$ as: \be
f(x)=\left\{ \begin{array}{ll} 1/x^2 & \mbox{for $x<-1$} \quad
  \mbox{(inflation)}\\ a_0+a_1 x+a_2 x^2+a_3 x^3 & \mbox{for
    $-1<x<x_0-1$} \quad \mbox{(transition)} \\ b_0(x+b_1)^2 &
  \mbox{for $x>x_0-1$} \quad \mbox{(radiation)},
   \end{array}
              \right.
\label{eq:smooth_radiation}
 \ee
where the 6 parameters $a_i$ and
$b_i$ can be fixed by imposing continuity of $f(x)$,
$f(x)^{\prime}$ and $f(x)^{\prime\prime}$ in the two points $x=-1$
and $x=x_0-1$. Since the corresponding expressions we find are
quite involved and they differ from the ones given in \cite{ford},
we give them in the Appendix.

\medskip

Let us now apply such a  parametrization to the different case when inflation is followed by a
period of kination. In this case one has
$\rho\propto 1/a^6$ and $H\propto 1/a^3$, so that $R\propto 1/a^6$ and
the adiabatic condition, $R \to 0$ is verified only asymptotically for
$\eta\rightarrow \infty$ and $a(\eta)\rightarrow\infty$. Recalling
$H=a^{\prime}/a^2$ one finds that $a^2(\eta)$ is a linear function of $\eta$
during kination.
The equivalent of
Eq.~(\ref{eq:smooth_radiation}) for a transition between inflation and
kination is then given by
\be
f(x)=\left\{ \begin{array}{ll}
                 1/x^2 & \mbox{for $x<-1$} \quad \mbox{(inflation)}\\
                a_0+a_1 x+a_2 x^2+a_3 x^3     & \mbox{for $-1<x<x_0-1$}
                \quad \mbox{(transition)}\\
                b_0+b_1 x     & \mbox{for $x>x_0-1$}\quad \mbox{(kination),}
                 \end{array}
              \right.
\label{eq:smooth_kination}
\ee
in which the 6 parameters $a_i$ and
$b_i$ can be fixed in the same way as in Eq.(\ref{eq:smooth_radiation}).
In this case the corresponding expressions are particularly
simple, and given by:
\be
\left\{ \begin{array}{l}
                 a_0=-\frac{1}{x_0}+6,\quad
                 a_1= -\frac{3}{x_0}+8,\quad
                 a_2= -\frac{3}{x_0}+3,\quad
                 a_3= -\frac{1}{x_0}\\
                 b_0=3 + 3x_0-x_0^2,\quad
         b_1=2+3x_0. \,\,\,\,\,\,\,\,\,
                 \end{array}
              \right.
\label{eq:a_b_coeff}
\ee

\begin{figure}
\begin{center}
\includegraphics[width=5.9cm]{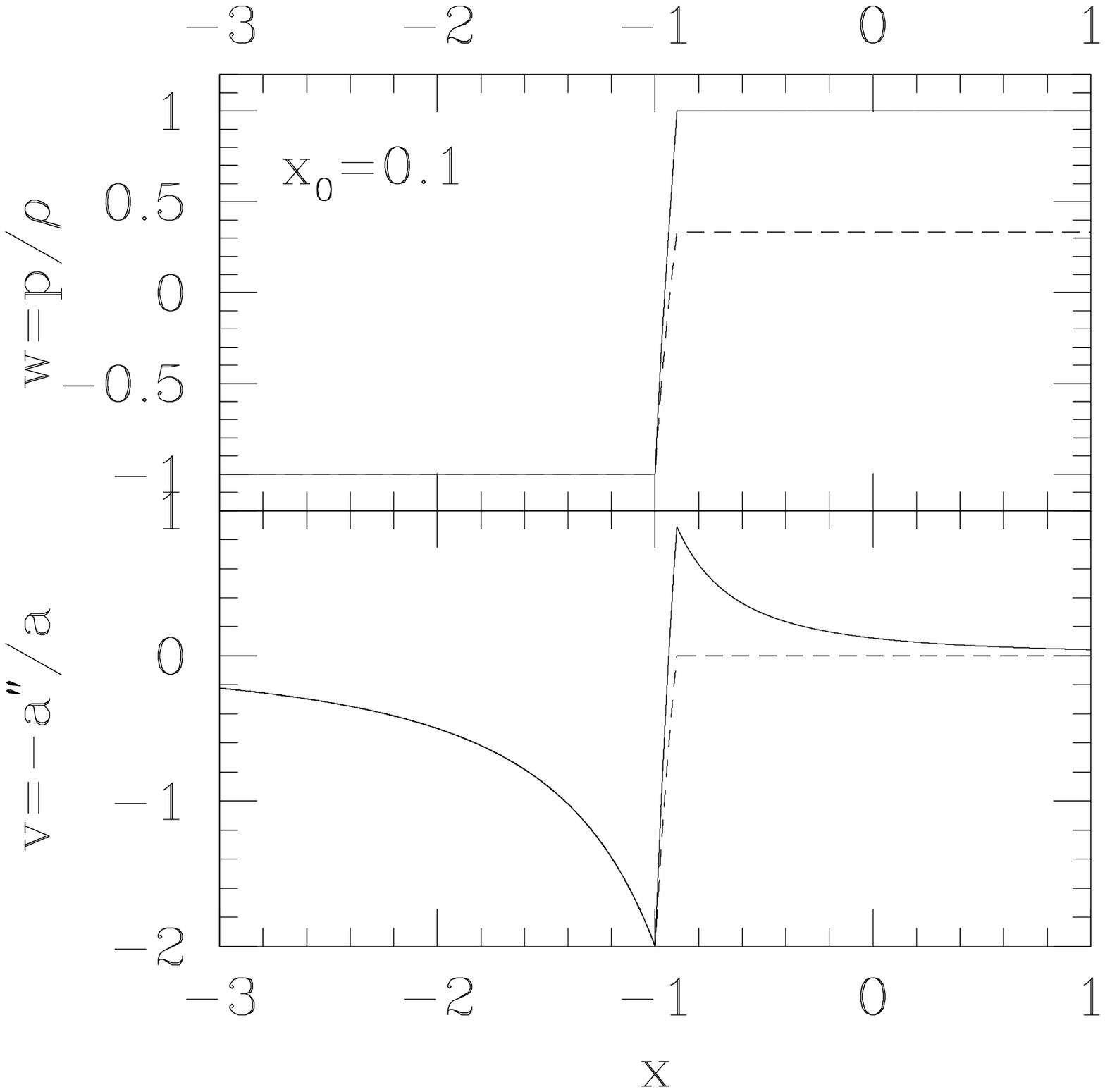}
\includegraphics[width=5.9cm]{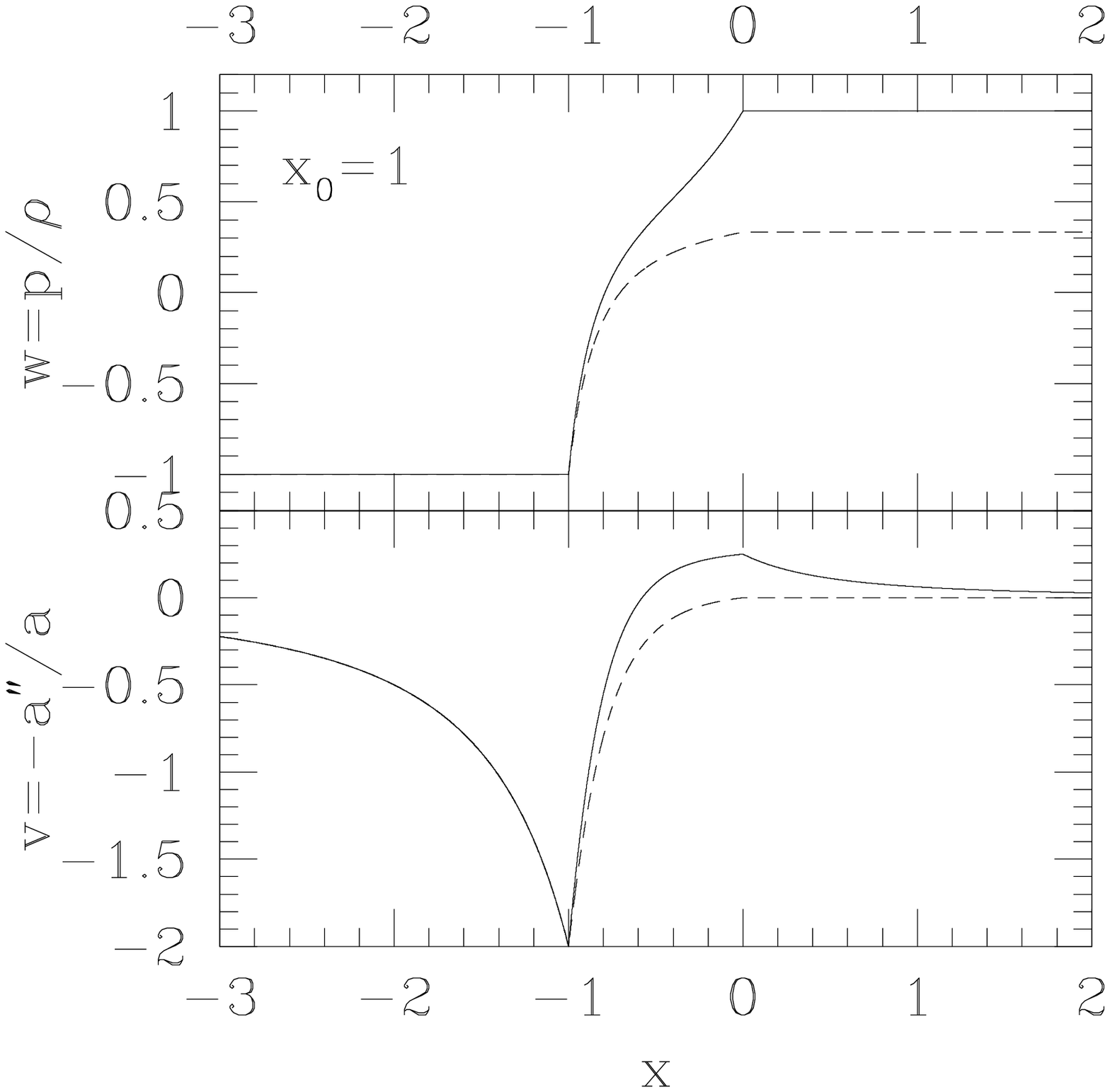}
\includegraphics[width=5.9cm]{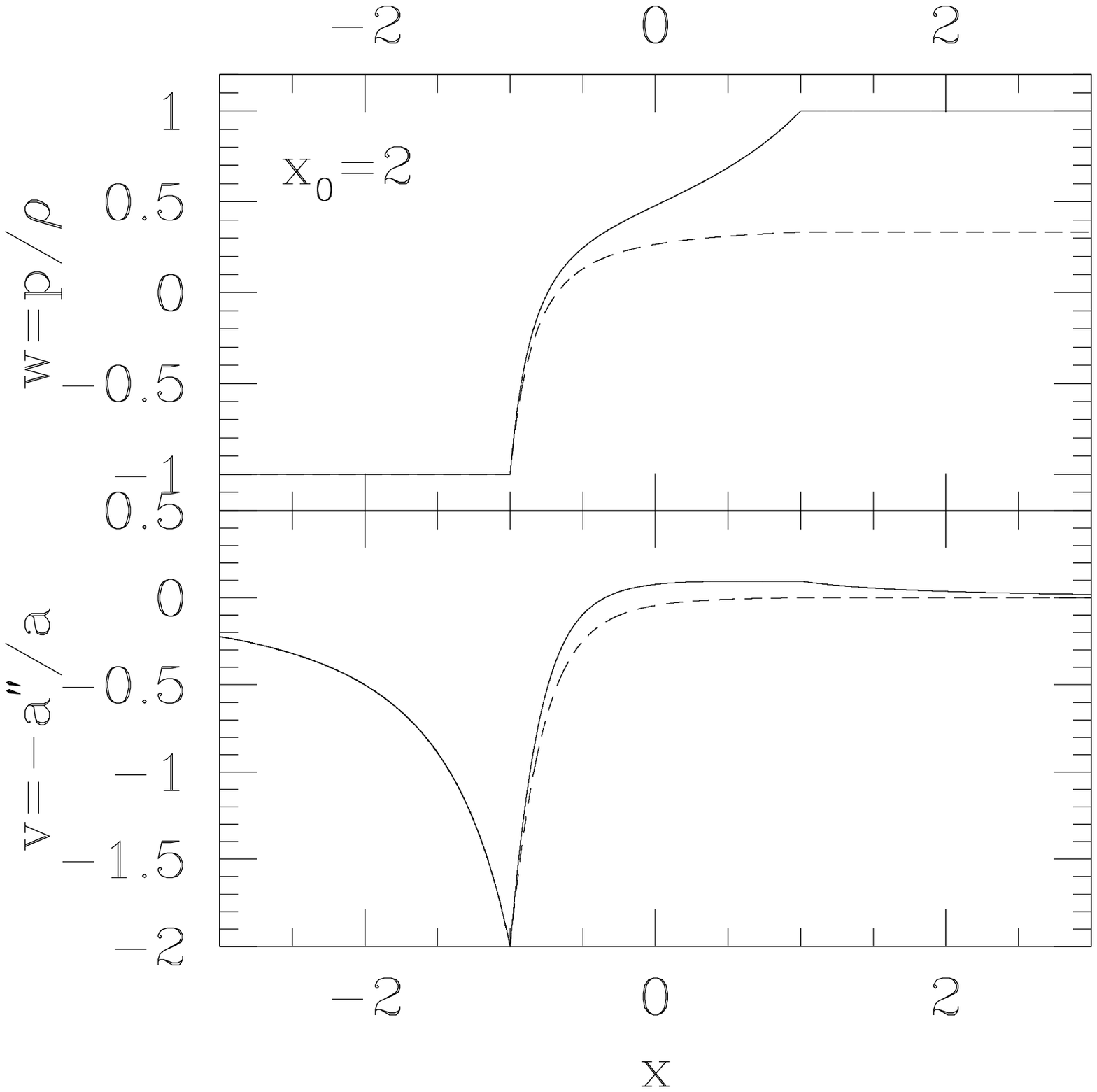}
\end{center}
\caption{Equation of state $w=p/\rho$ (upper panels) and function
  $V=-a^{\prime\prime}/a$ from Eq.~(\protect\ref{eq:omega_k}) (lower
  panels) as a function of $x_0=H_0\Delta\eta$, for $x_0=0.1$
  (left), $x_0=1$ (center) and $x_0=2$ (right).  Both cases when the
  Universe at the end of inflation is dominated by radiation (dashed
  line) and kination (solid line) are shown for comparison. The
  parametrizations of Eqs.~(\protect\ref{eq:smooth_radiation},
  \protect\ref{eq:smooth_kination}) are used.
\label{fig:V}}
\end{figure}

\begin{figure}
\begin{center}
\includegraphics[width=5.9cm]{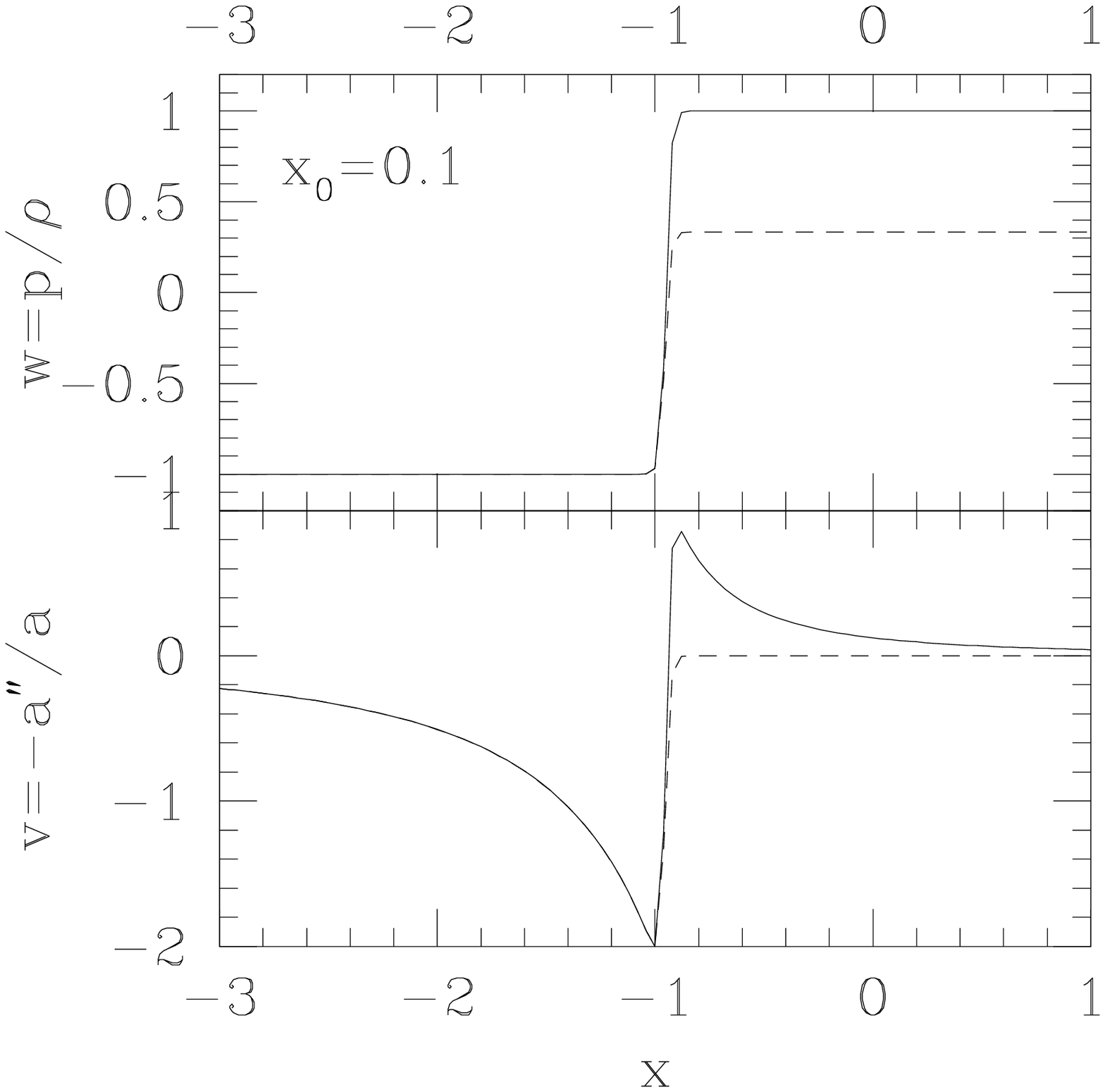}
\includegraphics[width=5.9cm]{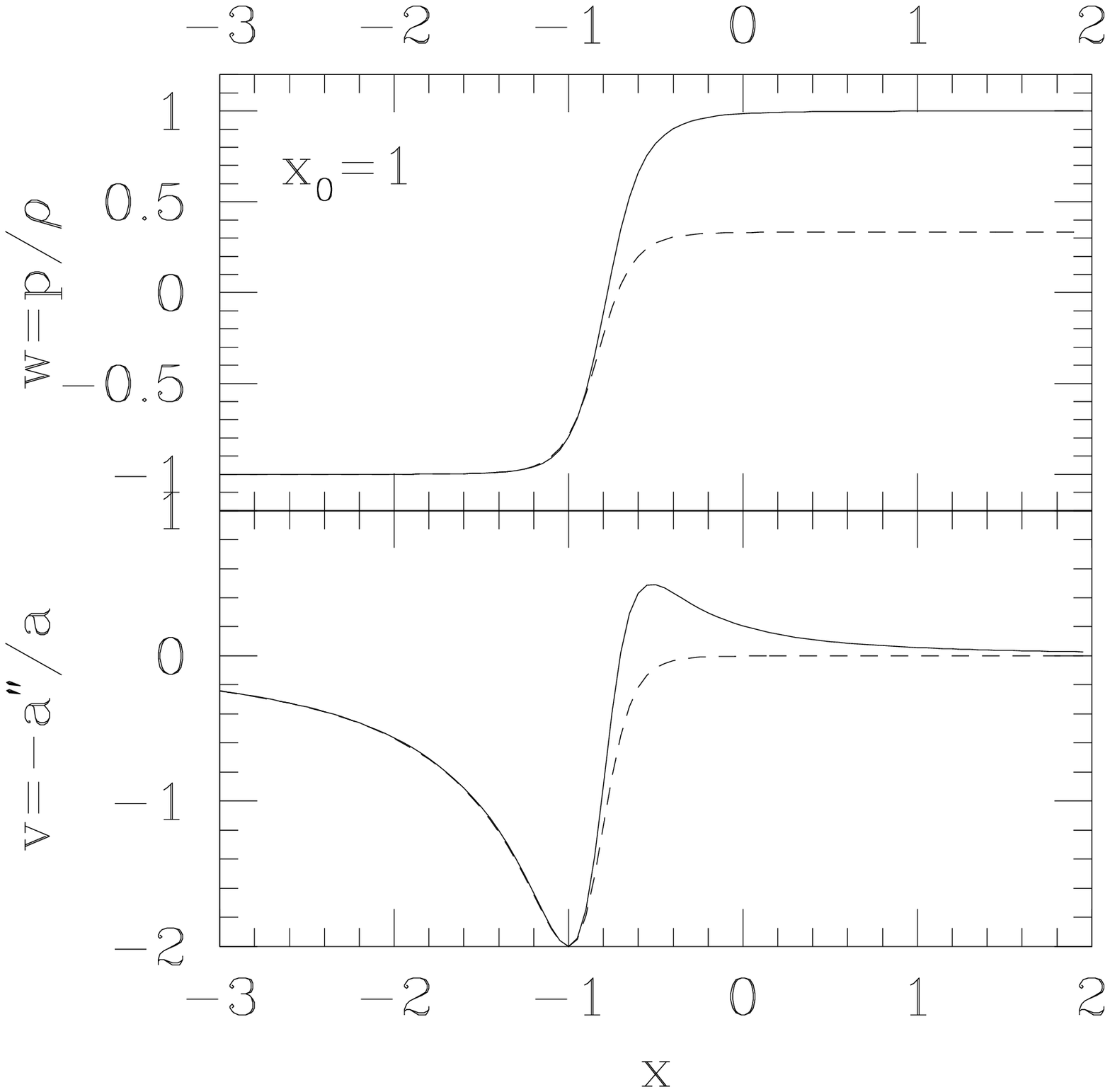}
\includegraphics[width=5.9cm]{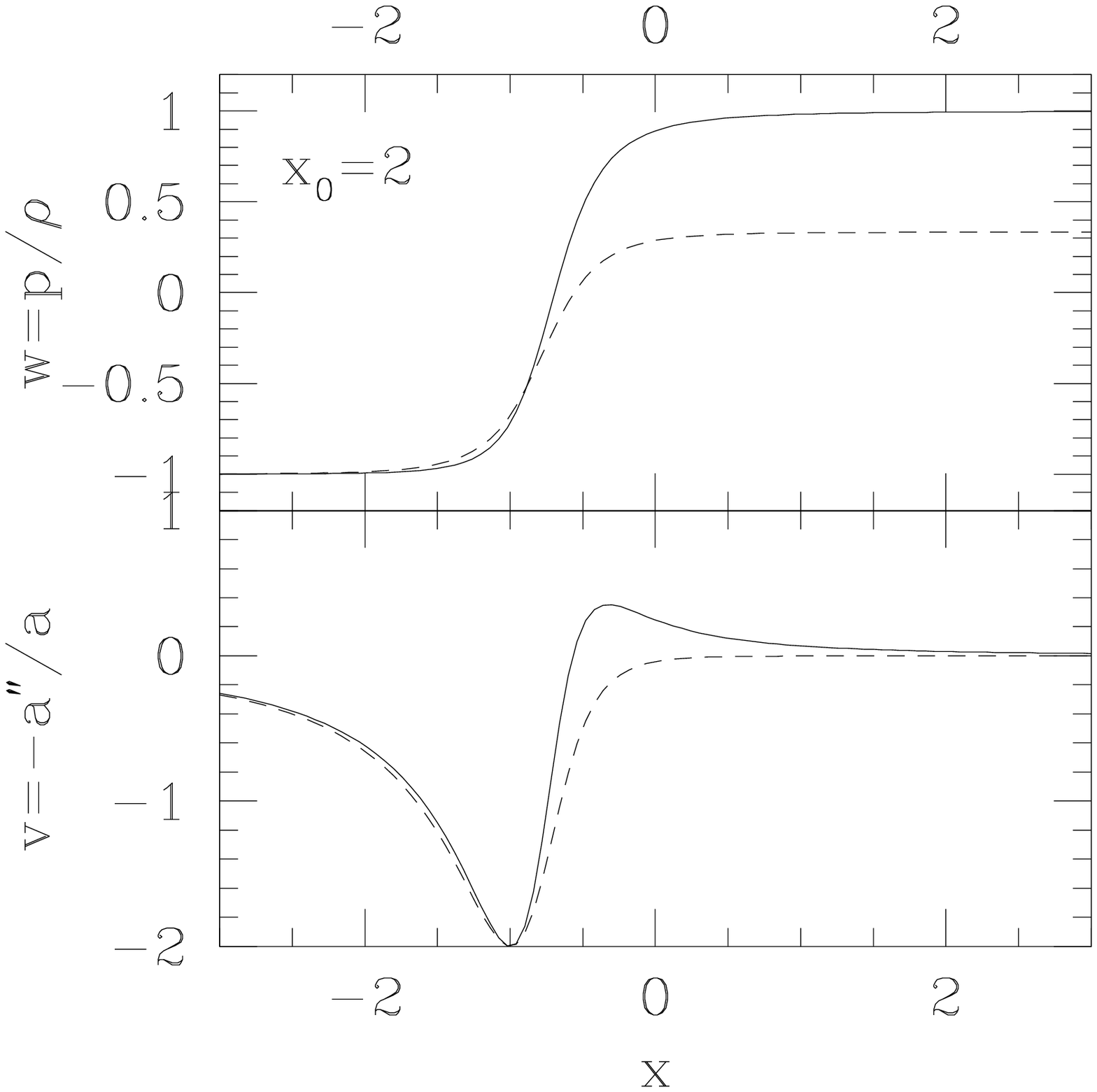}
\end{center}
\caption{The same as in Fig.\protect\ref{fig:V}, but using the parametrization of Eq. (\ref{eq:w}).
\label{fig:V_from_w}}
\end{figure}

In Fig \ref{fig:V} we plot the equation of state $w=p/\rho$ (upper
panel) and the function $V(x)$ from Eq. (\ref{eq:omega_k}) (lower
panel) for both cases of radiation domination (dashed line) and
kination domination (solid line) for $x_0=0.1$ (left), $x_0=1$
(center), $x_0=2$ (right), using the parametrizations of
Eqs.~(\ref{eq:smooth_radiation},\ref{eq:smooth_kination}). Taking
into account the approximate expression of the Bogolubov
coefficient given in Eq.~(\ref{eq:fourier}) it is possible from
this figure to anticipate that an enhancement of the process of
particle creation is expected whenever $x_0 \ll 1$, and especially
for kination compared to radiation. On the other hand, at large
$x_0$ the process is generally suppressed compared to the previous
case and is almost independent on $x_0$. Notice however that the
parametrization
(\ref{eq:smooth_radiation},\ref{eq:smooth_kination}) can be
considered as a reasonable choice capturing model-independent
properties of a generic transition only when $x_0 \ll 1$. When
$x_0\gsim 1$ the actual behavior of the functions $w(x)$ and
$V(x)$ will depend on the details of the model-dependent
transition.  For instance, the peculiar behavior of the function
$w(x)$ for the inflation--kination transition in the third panel
of Fig.~\ref{fig:V} is to be ascribed to an artifact of the
parametrization used rather than to the physical properties of a
realistic system.

\medskip

In order to estimate the model dependence of our results, in the following analysis we
will also adopt a smoother transition applied to the equation of state:
\beqarr
w(y)&=&\frac{w_f-1}{2}+\frac{w_f+1}{2}\tanh\left (\frac{2y}{y_0}\right ),
\label{eq:w}
\eeqarr
where $y\equiv \log(a)+\mbox{const}$, and
 $y_0$ parametrizes the duration of the transition in
e--foldings, while $w_f=1/3,1$ is the asymptotic equation of state at
late times for radiation and kination, respectively.
With the above
parametrization, the relation
 \be
\frac{d\log H}{d y}=\frac{d \log H}{d\log a}=-\frac{3}{2}(1+w),
\label{eq:dh}
 \ee
 can be integrated analytically to yield:
 \be
\tilde{H}\equiv H/H_0=e^{-\frac{3}{4}(1+w_f)\left [y+\frac{y_0}{2}
\left (\log \cosh\frac{2y}{y_0}+\log 2 \right )\right ]},
\label{eq:h_tilde}
 \ee
with $H_0=H(y=-\infty)$. The relation between $y$ and $x$ is
given by $dx/dy=1/(a\tilde{H})$. Identifying the transition
period with the interval $-y_0<y<+y_0$ ($w(\pm y_0)=\pm 0.96)$,
we have:
\be
x_0=\int_{-y_0}^{y_0}\frac{dy^{\prime}}{\tilde{H}(y^{\prime})a(y^{\prime})}.
\label{eq:x0}
\ee
 The quantities $w=p/\rho$ (upper panels) and $V(x)$
 (lower panels) in this parametrization are shown in Fig.~\ref{fig:V_from_w} for $x_0=0.1,1,2$,
 where $V(x)$ is obtained from the general relation:
 \be
 V(x)= -\frac{a^{\prime\prime}}{a}=-\frac{a(y)^2 \tilde{H}(y)
^2}{2}(1-3w(y)).
\label{eq:potential}
 \ee In Fig.~\ref{fig:V_from_w} we have chosen the (arbitrary)
 normalization of the scale factor $a \propto e^{y}$ and the origin of
 the conformal time variable $x$ in such a way that the minimum of
 $V(x)$ is for $x=-1$, and $V(x=-1)=-2$, in order to have a direct
 comparison with the previous parametrization discussed in
 Fig. \ref{fig:V}. That is, our normalization in terms of the variable
 $y$ is set to satisfy $V(y_{min})=-2$ and $x(y_{min})=-1$ where
 $y_{min}$ can be calculated by solving $dV/dy|_{y=y_{min}}=0$.
%
From Eq.~(\ref{eq:h_tilde}) the asymptotic behavior of the Hubble
constant follows in a straightforward way:
 \be
 \tilde{H}(y\gg+y_0)\simeq
e^{-\frac{3}{2}(1+w_f) y}= \frac{\left
[a(y_{min})e^{-y_{min}}\right
]^{\frac{3}{2}(1+w_f)}}{a^{\frac{3}{2}(1+w_f)}},
 \ee
which gives, for the case of kination,
 \beqarr \label{eq:b1_kination}
&&\tilde{H}(y\gg+y_0)\simeq \frac{b_1}{2 a^3}\quad\mbox{with}\quad
b_1=2 a(y_{min})^3 e^{-3 y_{min}}  .
 \eeqarr
 All the quantities discussed above can be expressed explicitly in terms of $y_0$.
In the case of a transition from inflation to kination ($w_f=1$),
we get:
 \beqarr
&&y_{min}=-\frac{y_0}{4}\log \left (\frac{(9
y_0+6)^{\frac{1}{2}}+(y_0+6)^{\frac{1}{2}}}{(9
y_0+6)^{\frac{1}{2}}-(y_0+6)^{\frac{1}{2}}}
\right),\label{eq:a_y_min}
 \eeqarr
 from which explicit forms of other variables can be derived.

Note that in the limit $y_0\rightarrow 0$ a power expansion allows
to recover the parametrization of Eq.~(\ref{eq:smooth_kination}).
In this case $a(y_{min}),\tilde{H}(y_{min})\rightarrow 1$, but
when making a direct comparison with the parametrization of
Eq.~(\ref{eq:smooth_kination}) the correspondence between $y_0$ and
$x_0$ is somewhat arbitrary since $w=\pm 1$ only when
$y\rightarrow \pm \infty$. However, by choosing the boundaries of
the transition at $y=\pm \bar{y}_0\equiv \pm c y_0$ in such a way
that $y_{min}\rightarrow -\bar{y}_0\simeq -x_0/2$ (so that in both
parametrizations the beginning of the transition coincides
asymptotically to the minimum of the potential), we find
 \be
 b_1 \rightarrow 2
e^{\frac{3}{2}x_0}\simeq 2+3 x_0 ,
 \ee
 which corresponds to the expression for $b_1$ found in the
last line of Eq.~(\ref{eq:a_b_coeff}) (The above
procedure corresponds to setting $c=\log(y_0)/4$)). With such
normalizations Figs.~\ref{fig:V} and \ref{fig:V_from_w} can be
directly compared.  As expected they differ when $x_0\gsim 1$
while they coincide when $x_0\ll 1$. Notice however that the
definition of the conformal time depends on the arbitrary
normalization of the scale factor $a$. For this reason in the
following section we will discuss our results expressing the
duration of the transition $x_0$ in term of the physical time $t$.
Of course the physical results will not depend on these
conventions.

As we will see later, $b_1$ is one of the important parameters
determining the temperature of the radiation component in
quintessential inflation models.  In the limit of an abrupt
transition ($\Delta t \to 0$), we have $b_1 \approx 2$ as
discussed above. In the limit of a slow transition, $b_1$ behaves
like: \be b_1 \propto (H_0\Delta t)^c, \ee where $\Delta t $ is
the transition time and the exponent $c$ is a model-dependent
parameter.  We find $c=0.5$ and $c\approx 0.48$ in the case of the
polynomial and {\it tanh} parametrization, respectively.  For the
latter, we define $\Delta t$ by $ \Delta t \equiv
\int^{y_0}_{-y_0}dy/H$.

\section{Numerical Results}
\label{sec:discussion}

\begin{figure}
\begin{center}
\includegraphics[width=8cm]{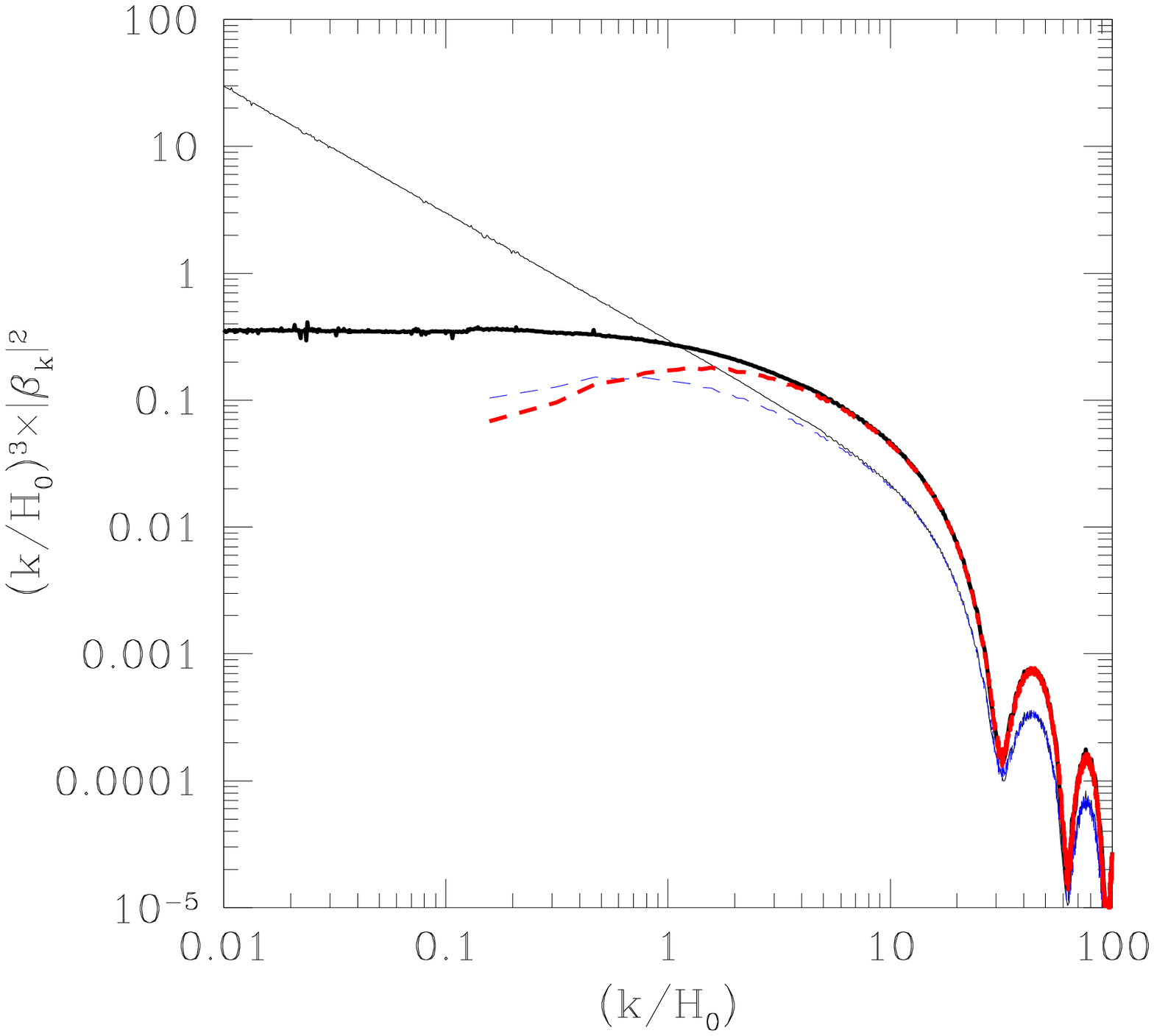}
\includegraphics[width=8cm]{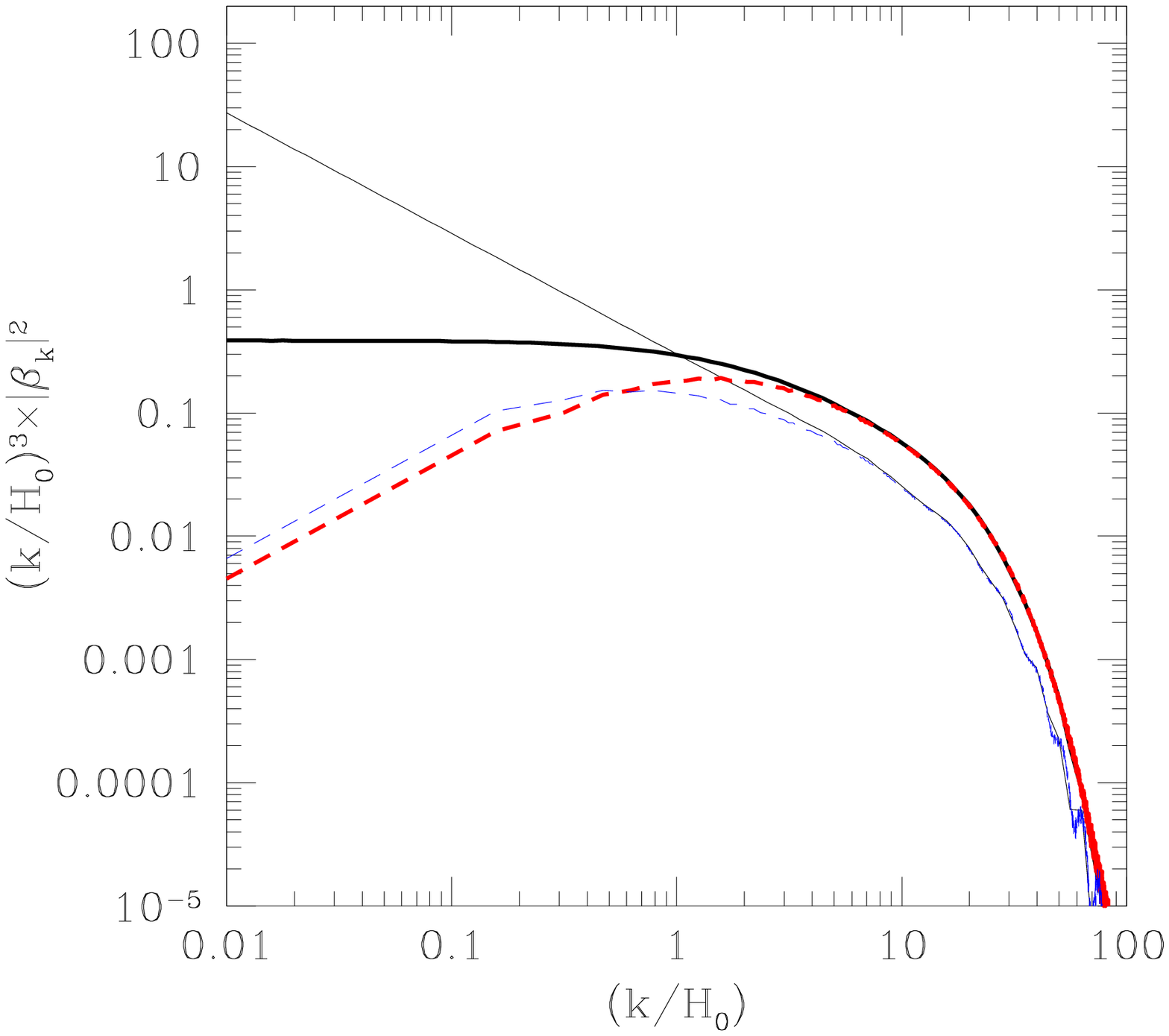}
\end{center}
\caption{$k^3\times |\beta_k|^2$ as a function of $k$ normalized to $H_0$
 in the case of inflation--radiation transition
  (thin lines) and inflation--kination transition (bold lines), and
  for $x_0=0.1$. Solid curves are the result of numerical integration
  of Eq.~(\protect\ref{eq:final_eq}), while dashed lines are obtained
  by making use of the perturbative approximation given in
  Eq.~(\protect\ref{eq:fourier}). In the left--hand panel the
  transition is modeled according to
  Eqs.~(\protect\ref{eq:smooth_radiation},\protect\ref{eq:smooth_kination}),
  while in the right--hand one the parametrization of
  Eq.~(\protect\ref{eq:w}) is used.
\label{fig:beta_k}}
\end{figure}

In this section we discuss our results for the amount of radiation
(\ref{eq:energy_density}) with the Bogolubov coefficient
$|\beta_k|^2$ given by the asymptotic value of
Eq.~(\ref{eq:final_beta}) at $\eta\rightarrow \infty$. The latter
is obtained in a straightforward way by solving numerically the
second-order differential equation (\ref{eq:final_eq}) with the
boundary conditions given in (\ref{eq:boundary_r}). In
Fig.~\ref{fig:beta_k} we present the quantity $k^3\times
|\beta_k|^2$, which enters the energy integral of
Eq.~(\ref{eq:energy_density}), as a function of $k$ for $x_0=0.1$
in the case of the inflation--radiation transition (thin lines)
and inflation--kination transition (bold lines). In the left-hand
panel the transition is parametrized according to
Eqs.~(\ref{eq:smooth_radiation},\ref{eq:smooth_kination}), while
in the right-hand one Eq.~(\ref{eq:w}) is used.  In both figures
solid lines are the result of numerical integration of
Eq.~(\ref{eq:final_eq}), while dashed lines are obtained by making
use of the perturbative approximation given in
Eq.~(\ref{eq:fourier}).  A couple of comments  are in order here.

\medskip

Fig.~\ref{fig:beta_k} shows an infrared divergence  for the
inflation-radiation transition, but not for the
inflation-kination transition.  Such a property can be studied
analytically in the limit of an instantaneous transition for which
we can obtain simple solutions for particle creation.  We assume
that the inflation-radiation transition occurs at $x=H_0 \eta =-1$
($x_0\equiv0$).  Apparently natural choices of the in-modes are:
 \begin{equation} \label{eq:modes_inflation}
 \chi_k^{in} = \left( {i\over k\eta}
 -1\right) {e^{-i k\eta} \over\sqrt{2 k}},
 \end{equation}
which are the solutions of Eq.~(2) during inflation. Now using the
out-modes during radiation in Eq.~(6), it is easy to  find
 \begin{equation}
 \left| \beta_k \right|^2 = {1\over 4 k^4},
 \end{equation}
which results in the infrared divergence of the energy density
(\ref{eq:energy_density}) for the modes $k\to 0$.  This implies
that the in-modes for $k\to 0$ correspond to unphysical states
since otherwise this large infrared contribution to the energy
density cannot represent a self-consistent solution to the
Einstein's equations. One way forward to solve this problem is to
find a suitable choice of the initial vacuum state such that the
two-point function for the field does not diverge in the infrared
limit \cite{ford77}. Once such choices are made to obtain finite
results for $k\to 0$, their contribution to the integration of the
total energy density becomes negligible. In this paper, we 
take a practical approach to cut off the  $k< H_0$ modes to
calculate the energy density. A possible alternative approach to
solve this problem would be to calculate the radiation energy
density by using the renormalized energy momentum tensor with a
proper regularization scheme \cite{birrell}, which would not
display this infrared behavior during the quasi de Sitter stage
\cite{iz}.

It is interesting to note that the situation is different for the
inflation-kination transition. Again assuming an instantaneous
transition at $x=-1$, one gets $V=H_0^2/(3+2x)^2$. While we can
take the in-modes as in Eq.~(\ref{eq:modes_inflation}) during
inflation, the out-modes during kination can be written as
 \begin{equation}
  \chi_k^{out} = {\sqrt{\pi}\over 2}
  \sqrt{z  \over H_0 } \left[
  \alpha_k H_0^{(2)}(\kappa z ) +
  \beta_k H_0^{(1)}(\kappa z) \right],
  \end{equation}
where $\kappa\equiv k/H_0$, $z\equiv x+ 3/2$ and $H_0^{(1,2)}$ are
the Hankel functions.  Matching the in-modes and out-modes at
$x=-1$ one readily finds for $k\to 0$,
 \begin{equation}
  \left| \beta_k \right|^2 \sim {1\over \pi k^3},
  \end{equation}
which is consistent with our numerical results in Fig.~3.  Therefore,
we can conclude that the radiation-kination transition may be free
from infrared divergences in reasonable models for the transition.

Another interesting observation is that the epoch of particle
production is confined to a short interval close to the minimum of the
potential $V(x)$, which, by our convention, is for $x=-1$ or
$y=y_{min}$.  This can be seen by considering the adiabaticity
parameter $\omega^{\prime}/\omega^2$ which becomes much smaller than 1
for the adiabatic regime.  In the case of a slow transition ($x_0 \gg
1$) the adiabaticity parameter is suppressed at all times except for
$x\simeq -1$, so particle creation is a quasi--instantaneous process
even in this case, as in the situation of an abrupt transition with
$x_0\ll 1$. For this reason when the transition time is large the
amount of particle creation becomes a very slowly varying function of
$x_0$ or $H_0 \Delta t$, as will be shown below.

\medskip

\begin{figure}
\begin{center}
\includegraphics[width=8cm]{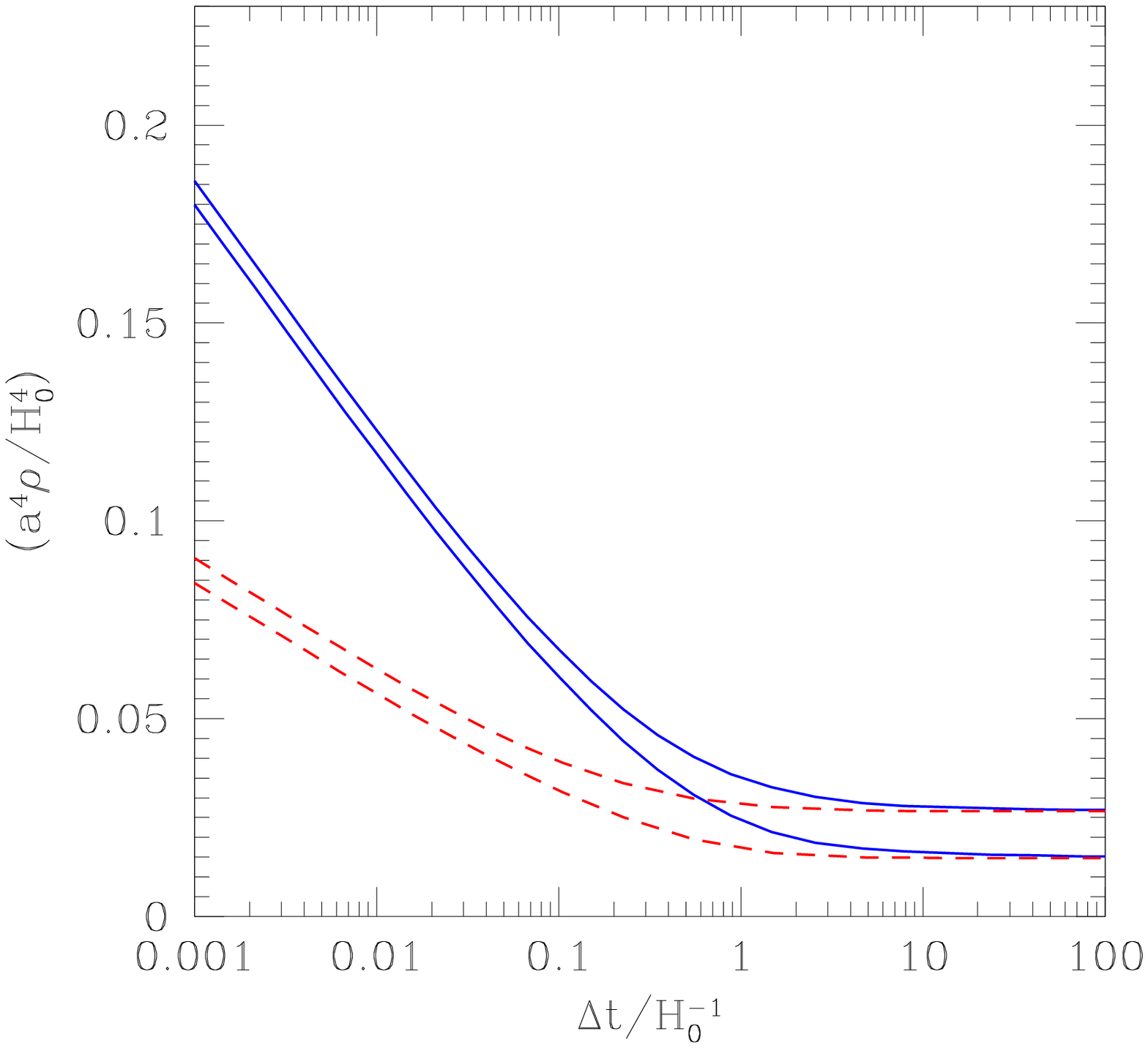}
\includegraphics[width=8cm]{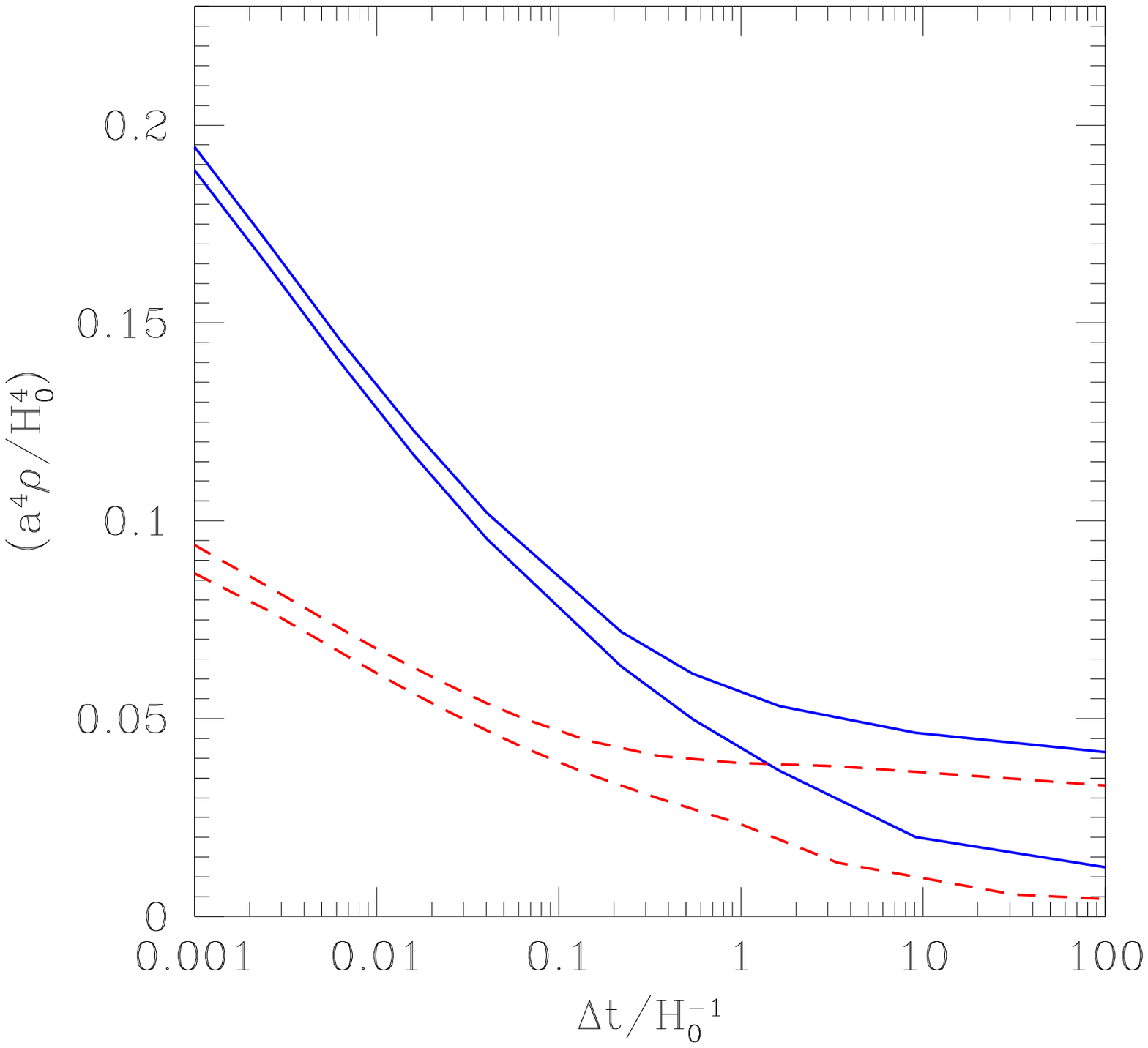}
\end{center}
\caption{The energy density per scalar degree of freedom produced by
  gravitational reheating (normalized to $H_0^4$) as a function of the
  duration in physical time of the transition (normalized to
  $1/H_0$). Solid lines are for the inflation--kination transition and
  dashed
  lines are for inflation--radiation transition.  In both cases the upper line
  is the result of a numerical integration of
  Eq.~(\protect\ref{eq:final_eq}) and the lower one is obtained by
  making use of the approximation of Eq.~(\protect\ref{eq:fourier}).
  In the left--hand panel the transition is modeled according to
  Eqs.~(\protect\ref{eq:smooth_radiation},\protect\ref{eq:smooth_kination}),
  while in the right--hand one the parametrization of
  Eq.~(\protect\ref{eq:w}) is used.
\label{fig:creation}}
\end{figure}

The computation of the energy density is shown in
Fig.~\ref{fig:creation}, where the quantity $a^4 \rho$ normalized
to $H_0^4$, is plotted as a function of $H_0 \Delta t$ in both
cases of a transition between inflation and radiation (dashed
lines) and inflation and kination (solid lines). Notice how, as
previously anticipated, the comoving energy density $a^4\rho$
shows a plateau at large $\Delta t$, while an enhancement is
present at smaller $\Delta t$. In the case of kination, due to the
fact that the function $V$ has a second peak at positive $\eta$
for small $\Delta t$ (see the lower part of the left panel of
Fig.\ref{fig:V}), which is not present for radiation due to a
vanishing Ricci scalar $R$, the quantity $a^4\rho$ is larger by
about a factor of two compared to the case of radiation, and in
both cases, we find the behavior of ultra-violet divergence:
$a^4\rho \propto \ln(H_0\Delta t)$ for $\Delta t \to 0$ as
discussed previously.  However, the amount of radiation produced
by gravitational reheating for large $\Delta t$ is similar in the
two cases, since in that case the potential $V$ for kination is
strongly suppressed for $\eta>0$ (see the lower part of the
right--hand panel in Fig.~\ref{fig:V}) and thus very similar to
the case of radiation. From Fig.~\ref{fig:creation} we obtain the
typical amount of particle creation parameterized by $I\equiv
a^4\rho/H_0^4$:
 \beqarr && 0.03\lsim I \lsim 0.08 \;\;\;{\rm
Inflation \to Radiation} \nonumber
\\ && 0.03\lsim I \lsim 0.19 \;\;\;{\rm Inflation \to Kination},
\label{eq:I_intervals}
 \eeqarr
restricting ourselves to $H_0\Delta t\gsim 10^{-3}$.

\section{Radiation--kination equality}
\label{sec:equality}

A crucial element for the phenomenology of quintessential inflation is
the epoch when the standard cosmology with radiation domination
starts. The amount of radiation produced by gravitational reheating is
very small, so that at the end of inflation the energy of the Universe
is dominated by the kinetic energy of the quintessence field.
However, the energy density of kination $\rho_K$ is red-shifted away
much faster than that of radiation $\rho_R$ as \be \rho_{K}\propto
a^{-6} \quad\mbox{and}\quad \rho_{R}\propto a^{-4},\label{eq:a_dep}
\ee and thus the latter can eventually dominate at some temperature
much below the initial (reheat) temperature $T_{I}$. Let us define the
kination--radiation equality temperature $T_*$ by
 \be
\rho_{K}(T_*)=\rho_{R}(T_*), \label{eq:tstar}
 \ee
which should follow
the constraint: $T_* \gsim$ 1 MeV, in order not to spoil
nucleosynthesis.

Consider thermalization of the gravitationally produced
particles  occurring  at $a_I$ with the corresponding (reheat)
temperature $T_I$.
 At this moment, the radiation
energy density can be expressed as
 \be
\rho_R^I=\frac{g_I I H_0^4}{a_I^4}=\frac{\pi^2}{30} g_I  T_I^4,
\label{eq:rho_r}
 \ee
from which one determines $a_I T_I$ given $I$ and $H_0$.
The specific value of $T_I$   can be explicitly
calculated if the interaction rates of particles are given in a particular
model \cite{chung07}.  For instance,  assuming a typical interaction rate of
$\Gamma \sim \alpha^2 T$ with a certain coupling strength $\alpha$, and
equating it with the
Hubble parameter $H=b_1/2a^3$ during kination at $a_I$ or $T_I$, we find
 \be
  T_I = \alpha \left( 30 \over \pi^2\right)^{3\over8}
  \left( 2 I^{3\over4} \over b_1\right)^{1\over2} H_0,
 \ee
which can be much smaller than the Gibbons-Hawking temperature $H_0/2\pi$ \cite{hawking}.
Irrespectively of the specific values of $a_I$ or $T_I$, we can determine the temperature $T_*$ when
the radiation domination starts.  Using Eq.~(\ref{eq:rho_r}) and the relation of iso-entropic
expansion, $
 g_I\, a_I^3 T_I^3 = g_* \, a_*^3 T_*^3 $,
 one obtains
 \be
 a_* T_* = \left( g_I \over g_*\right)^{1\over3} \left( 30 I \over \pi^2\right)^{1\over4} H_0.
 \label{eq:aTstar}
 \ee
 Since the energy densities of radiation  $\rho^*_R$ and
 kination $\rho^*_K$ at $a_*$ or $T_*$ are
  \beqarr
 && \rho^*_R = {\pi^3\over 30}g_* T_*^4 \nonumber \\
 && \rho^*_K = 3 m_P^2 H_0^2 {b_1^2 \over 4 a_*^6},
 \label{eq:rho_r_star}
  \eeqarr
the conditions (\ref{eq:tstar}) and (\ref{eq:aTstar})
enable us to get
 \be
  T_*
=\frac{2\times 30^{\frac{1}{4}}}{\sqrt{3\pi}}\,
    \frac{g_I}{g_*^{\frac{1}{2}}}\,
    \frac{I^{\frac{3}{4}}}{b_1}\,
     \frac{H_0^2}{m_P}.
\label{eq:tstar_final}
  \ee
In the above equation $b_1$ is given by Eq.~(\ref{eq:a_b_coeff})
or (\ref{eq:b1_kination}), depending on which
parametrization for the transition is adopted. Moreover, notice
that both quantities $I$ and $b_1$ depend on the arbitrary
normalization of the scale factor, while, as expected,
in Eq.~(\ref{eq:tstar_final}) the combination $I^{3/4}/b_1$ entering the
physical quantity $T_*$ or $T_I$ does not.

\begin{figure}
\begin{center}
\includegraphics[width=8cm]{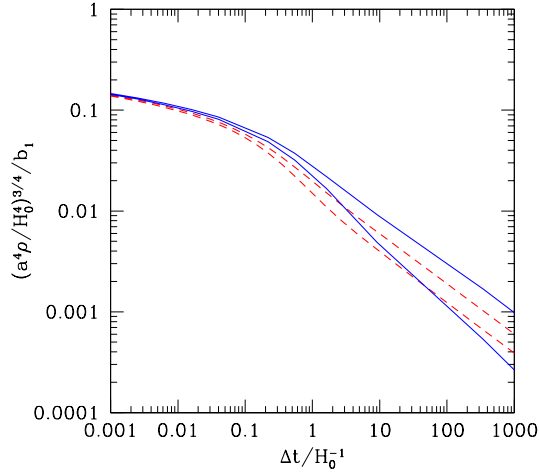}
\end{center}
\caption{The combination $I^{\frac{3}{4}}/b_1$, entering into the
  kination--radiation equality temperature $T_*$ given in
  Eq.(\protect\ref{eq:tstar_final}), as a function of the transition
  time $\Delta t/H_0^{-1}$ expressed in absolute time.  Solid
  lines are for the
  transition given in Eq.~(\protect\ref{eq:w}), and  dashed
  lines are for the transition in
  Eq.~(\protect\ref{eq:smooth_kination}).
  In both cases the upper line is the result of a numerical
  integration of Eq.~(\protect\ref{eq:final_eq}) and the lower one is
  obtained by making use of the approximation of
  Eq.~(\protect\ref{eq:fourier}).
\label{fig:I_b1}}
\end{figure}

In Fig.~\ref{fig:I_b1} the combination $I^{3/4}/b_1$ is plotted in
terms of the transition time $\Delta t/H_0^{-1}$.  In the figure the
solid lines show $I^{3/4}/b_1$ when the transition is modeled
according to Eq.~(\protect\ref{eq:w}), while in the dashed lines the
transition is modeled according to
Eqs.~(\protect\ref{eq:smooth_radiation},\protect\ref{eq:smooth_kination}).
In both cases the upper line is the result of a numerical integration
of Eq.~(\protect\ref{eq:final_eq}) and the lower one is obtained by
making use of the approximation of Eq.~(\protect\ref{eq:fourier}).
Note that the quantity
$I^{3/4}/b_1$ shows the scaling behavior:
 \be
 {I^{3\over4} \over b_1} \sim
 \left\{ \begin{array}{ll}
      0.06\, [ \ln{1 \over H_0\Delta t} ]^{3\over4}
                & \mbox{for}\quad \Delta t\to 0 \\
    \mbox{$^{0.02}_{0.026}$} (H_0\Delta t)^{\mbox{\tiny{$^{-0.5}_{-0.48}$}}}
            & \mbox{for}\quad \Delta t\to \infty,
    \end{array}
 \right.
 \ee
 where the two values in the second line correspond to the
 transitions given in Eqs.~(16) and (18), respectively.  From this, we find the following
 numerical value of $T_*$:
  \be
  T_* \sim {g_I \over \sqrt{g_*}} {H_0^2 \over m_P}
   \left\{ \begin{array}{ll}
      0.09 \, [\ln{1 \over H_0\Delta t} ]^{3\over4}
                & \mbox{for}\quad \Delta t\to 0 \\
   \mbox{$^{0.03}_{0.039}$}\, (H_0\Delta t)^{\mbox{\tiny$^{-0.5}_{-0.48}$}}
            & \mbox{for}\quad \Delta t\to \infty,
    \end{array}
 \right.
 \label{eq:Tstar_value}
  \ee
 which depends two important model parameters: the inflation scale $H_0$ and the transition
 time $\Delta t$.

\section{Conclusions}
In this paper we have provided a detailed study of gravitational
reheating in quintessential inflation generalizing previous
analyses only available for the standard case when inflation is
followed by an era dominated by the energy density of radiation.
In the quintessential inflation scenario both inflation and dark
energy are caused by the same scalar field: initially the
inflaton/quintessence potential energy drives inflation, which is
then followed by a kination period when the energy density of the
Universe is dominated by the kinetic energy of the field; this
kination period is eventually ended by radiation domination, while
later on the potential energy of the quintessence field prevails
again, providing the dark energy observed today. In quintessential
inflation, among several ways to produce radiation and thus reheat
the Universe, gravitational particle creation is the minimal
scenario being able to provide a sufficient amount of radiation
in spite of a much lower efficiency compared to other mechanisms,
thanks to the fact that the kination energy density is
red--shifted at a much faster rate compared to radiation.  In this
scenario, the kination-radiation equality temperature $T_*$ is an
important quantity which has implications for various cosmological
events like baryogenesis, dark matter decoupling and
nucleosynthesis, etc.

Motivated by this, we performed a numerical analysis of Bobolubov
transformations to calculate the radiation energy density produced
gravitationally, and determine the dependence of $T_*$ on the
inflation scale $H_0$ and the transition time $\Delta t$ of
inflation to kination. To study how our conclusions depend on the
details in modelling the transition period we considered two
different parameterizations: a polynomial expansion of $a^2(\eta)$
and a parametrization of the equation of state making use of an
tangent hyperbolic function. Our main results are summarized in
Eq.~(\ref{eq:Tstar_value}).  In both cases we obtain numerically
similar results, both in the small and in the large $\Delta t$
limit which may hold for generic models with a reasonable
transition behavior.  It is also interesting to observe that the
inflation-kination transition does not show an infrared divergence
for natural choices of the in-modes which represented
unphysical vacuum states for small $k$ in the case of
the inflation-radiation transition.
Nevertheless the problem of such an infrared divergence
would be removed in a consistent way by using
the renormalized energy momentum tensor, which could provide an independent
study of the gravitational reheating process to confirm our results.
We leave this issue to a future investigation.

\bigskip

 {\bf ACKNOWLEDGEMENT}

 \noindent EJC thanks Daniel Chung
for useful discussions.

\appendix

\section{}
\label{sec:appendix}

We give here the explicit expressions for the parameters $a_i$ and
$b_i$ introduced in Eq.~(\ref{eq:smooth_radiation})
for the case of the inflation-radiation transition:
 \be
\left\{ \begin{array}{l}
                 a_1=3 a_0-10\\
                 a_2=3 a_0-15\\
                 a_3=a_0-6\\
                 b_0=3[a_0(1+t)-5-6t]\\
         b_1=\frac{a_1+2 a_2 t +3 a_3 t^2}{2 b_0}-t \,,
                 \end{array}
              \right.
 \ee
where $a_0$ is the physical solution of the quadratic
equation:
\be
C_1 a_0^2 +C_2 a_0 + C_3=0,
\label{eq:quadratic}
\ee
chosen in such a way that $f(x)=a^2(x)$ is always
positive.  In (\ref{eq:quadratic}) the constants $C_i$ are given by
\be
\left\{ \begin{array}{l}
                 C_1=K_1 K_7-K_3 K_5\\
                 C_2=K_1 K_8+K_2 K_7 -K_4 K_5 -K_3 K_6\\
                 C_3=K_2 K_8-K_6 K_4,
                 \end{array}
              \right.
\ee
where
\be
\left\{ \begin{array}{l}
                 K_1=3+6t+3t^2\\
                 K_2=-10-30 t -18 t^2\\
                 K_3=6+6 t\\
                 K_4=-30-15 t\\
                 K_5=2+6 t+6 t^2 +2 t^3\\
                 K_6=-20 t-30 t^2-12 t^3\\
         K_7=3+6 t+ 3 t^2\\
                 K_8-10-30 t -18 t^2,
                 \end{array}
              \right.
\ee
and $t\equiv x_0-1$. The above equations differ from
Eqs.~(18,19) in Ref.~\cite{ford}, which do not yield continuity
of $f^{\prime}(x)$ and
$f^{\prime\prime}(x)$ in $x=x_0-1$, as we checked by numerical
inspection.

\end{document}